\documentstyle[12pt]{article}
\def\al{\alpha}
\def\be{\beta}

\def\de{\delta}

\def\et{\eta}

\def\la{\lambda}

\def\De{\Delta}

\def\fr#1#2{{{#1} \over {#2}}}

\def\frac#1#2{{\textstyle{{#1}\over {#2}}}}

\def\lsim{\mathrel{\rlap{\lower4pt\hbox{\hskip1pt$\sim$}}
    \raise1pt\hbox{$<$}}}
\def\gsim{\mathrel{\rlap{\lower4pt\hbox{\hskip1pt$\sim$}}
    \raise1pt\hbox{$>$}}}
\def\sqr#1#2{{\vcenter{\vbox{\hrule height.#2pt
         \hbox{\vrule width.#2pt height#1pt \kern#1pt
         \vrule width.#2pt}
         \hrule height.#2pt}}}}

\newcommand{\beq}{\begin{equation}}
\newcommand{\eeq}{\end{equation}}
\newcommand{\bea}{\begin{eqnarray}}
\newcommand{\eea}{\end{eqnarray}}
\newcommand{\rf}[1]{(\ref{#1})}

\renewenvironment{thebibliography}[1]
 { \rm
   \begin{list}{\arabic{enumi}.}
    {\usecounter{enumi} \setlength{\parsep}{0pt}
     \setlength{\itemsep}{3pt} \settowidth{\labelwidth}{#1.}
     \sloppy
    }}{\end{list}}

\begin{document}
\titlepage

\begin{flushright}
{IUHET 281\\}
{hep-ph/9410325\\}
{June 1994\\}
\end{flushright}
\vglue 1cm

\begin{center}
{{\bf  QUANTUM DEFECTS AND THE LONG-TERM BEHAVIOR\\
	OF RADIAL RYDBERG WAVE PACKETS
\\}
\vglue 1.0cm
{Robert Bluhm$^a$ and V. Alan Kosteleck\'y$^b$\\}
\bigskip
{\it $^a$Physics Department\\}
\medskip
{\it Colby College\\}
\medskip
{\it Waterville, ME 04901, U.S.A\\}
\bigskip
{\it $^b$Physics Department\\}
\medskip
{\it Indiana University\\}
\medskip
{\it Bloomington, IN 47405, U.S.A.\\}

}
\vglue 0.8cm

\end{center}

{\rightskip=3pc\leftskip=3pc\noindent
\baselineskip=20pt
We show that a theoretical description of
radial Rydberg wave packets in alkali-metal atoms
based solely on hydrogenic wave functions and energies
is insufficient to explain data
that could be obtained in pump-probe experiments
with current technology.
The modifications to long-term
revival times induced by quantum defects
cannot be obtained by direct scaling of the hydrogenic results.
Moreover,
the effects of laser detuning and quantum defects are different.
An alternative approach providing analytical predictions
is presented.

}

\vskip 1truein
\centerline{\it To appear in the December 1994 issue of
Rapid Communications, Physical Review A}

\vfill
\newpage

\baselineskip=20pt

Radial Rydberg wave packets are produced when a short
laser pulse excites a coherent superposition of states
with no other fields present
\cite{ps,az}.
A single pulse produces a packet
with p-state angular distribution
but localized in the radial coordinate.
Initially,
the packet moves between the
apsidal points of a classical keplerian orbit.
The radial-coordinate uncertainty product $\De r\De p_r$
correspondingly oscillates between values
large compared to $\hbar$ and values close to minimum uncertainty.
Such oscillations are characteristic of a squeezed state.
Indeed,
a hydrogenic radial Rydberg wave packet
at its first pass through the outer apsidal point
may be modeled as a type of squeezed state,
called a radial squeezed state
\cite{rss}.
The motion of a radial squeezed
state undergoes a cycle involving collapse
and fractional/full revivals
characteristic of Rydberg wave packets
\cite{ps,az,ap,nau1,tenWolde,yeazell1,yeazell2,yeazell3,meacher}.

Experiments on single-electron radial Rydberg wave
packets typically involve alkali-metal atoms.
It turns out that radial Rydberg packets
in alkali-metal atoms can also be modeled using
radial squeezed states
\cite{susywave}.
In general,
the core electrons of an alkali-metal atom cause deviations
of the eigenenergies from hydrogenic values,
which can be characterized
by quantum defects
$\de (n,l)$
depending on the principal quantum number $n$
and the angular-momentum quantum number $l$.
For Rydberg states $n$ is large
and the $\de (n,l)$ attain
asymptotic values $\de (l)$ independent of $n$.
The energies in atomic units are then
$E_{n^\ast} = {-1}/{2 {n^{\ast 2}}}$,
where $n^\ast = n - \de (l)$.

Recently,
the revival structure of Rydberg wave
packets both in hydrogen and in alkali-metal atoms
has been studied
for times much greater than the full revival time $t_{\rm rev}$
\cite{sr,sr2,ftnt}.
On a time scale $t_{\rm sr} \gg t_{\rm rev}$,
a new cycle of fractional/full revivals commences.
At times $t_{\rm frac}$ equal to linear combinations
of $t_{\rm sr}$ and $t_{\rm rev}$,
the packet takes the form of a sum of
macroscopically distinct subsidiary packets.
In certain cases,
only one subsidiary packet forms,
more closely resembling the initial packet
than the full-revival one.
The motion of the packet at times near $t_{\rm frac}$
is periodic,
with period $T_{\rm frac}$
given as a linear combination of the revival time $t_{\rm rev}$
and the classical orbital period $T_{\rm cl}$.
The autocorrelation function exhibits peaks near
times $t_{\rm frac}$ with periodicities $T_{\rm frac}$.

In this paper,
we consider predictions for experimental results
from different theoretical descriptions
of the long-term evolution and revival structure of
a radial Rydberg wave packet in an alkali-metal atom.
In particular,
we examine the suitability of an analysis of the long-term behavior
based entirely on hydrogenic wave functions and energies,
commonly used to describe such a wave packet.
We show that this approach
is inadequate for a complete description
of phenomena accessible to
the present generation of pump-probe experiments.
For wave packets in alkali-metal atoms,
where quantum defects are present,
we also show that effects arising from laser detuning
and from quantum defects are different.

For definiteness,
we assume an experimental configuration
with pump-probe detection
involving either time-delayed photoionization
\cite{az}
or phase modulation
\cite{noord,broers,christian}.
In either case,
we take the packet to be produced by
single-photon excitation from the ground state,
yielding a p-state angular distribution.
The superposition of eigenstates resulting
from the initial laser pulse is characterized
by a distribution $c_{n^\ast}$.
The quantity $\vert c_{n^\ast} \vert^2$
describes the relative contributions of eigenstates with
energies $E_{n^\ast}$.
A short pulse produces a distribution of states with finite width
centered on a value $N^\ast$,
which need not correspond to an exact atomic resonance.

Both pump-probe methods produce an ionization signal
displaying periodicities at the revival times.
Ref.\ \cite{yeazell2} describes
the observation of the photoionization signal for
p-state radial wave packets in potassium with $N^\ast \simeq 65.2$,
for delay times up to $t \approx \fr 1 2 t_{\rm rev}$.
The resolution of the data is sufficient to observe
a relative phase shift approximately equal to
$\fr 1 2 T_{\rm cl}$ between the initial peaks
near $t \approx 0$ and the revival peaks
near $t \approx \fr 1 2 t_{\rm rev}$.
More recently,
fractional revivals up to seventh order have been
detected using the more sensitive phase-modulation technique
\cite{wals}.
These experiments,
ranging over times up to $\fr 1 2 t_{\rm rev}$,
resolve peaks in the ionization signal in rubidium with periods
as small as $\fr 1 7 T_{\rm cl}$ for $N^\ast \simeq 53.3$
and $\fr 1 4 T_{\rm cl}$ for $N^\ast \simeq 46.5$.

For present purposes,
we consider a hypothetical but feasible pump-probe experiment
with a delay line
such that the structure of the ionization signal
can be examined at times on the order of $t_{\rm sr}$.
Although the time scale $t_{\rm sr}$
is greater than $t_{\rm rev}$
for the range of $N^\ast$ values of interest
\cite{sr},
it is still several orders of magnitude smaller
than the lifetimes of the Rydberg states.
Since the periodicities in the peaks of the
ionization signal match the periodicities
of the underlying packet,
a comparison of different theoretical models
with each other and with experiment can be made
by direct study of the wave packet.

In what follows,
we take as experimental input to any given theory
the measured values of the times $t_{\rm frac}$
at which periodicities appear in the ionization signal,
the corresponding periods $T_{\rm frac}$,
and the central wavelength of the laser
used to excite the packet.
{}From the latter,
the mean value of the energy $E_{N^\ast}$
and hence $N^\ast = \sqrt{- 1/2 E_{N^\ast}}$
may be determined.
This mean energy may not correspond
to an atomic resonance.

The first model we consider
involves hydrogenic eigenenergies and eigenstates.
It is conceptually simple and is often used
to study Rydberg packets.
The wave function is expanded as
\beq
\Psi ({\vec r},t) = \sum_{n=1}^{\infty} c_n
\varphi_n ({\vec r}) \exp \left( -i E_n t \right)
\quad ,
\label{hpsit}
\eeq
where $\varphi_n ({\vec r})$ are p-state hydrogenic
wave functions,
$E_n = - 1/{2 n^2}$,
and the distribution $c_n$
has finite width and is centered around $N^\ast$.
The superposition \rf{hpsit}
therefore has mean energy matching that of the mean
energy of the packet.
\it A priori, \rm
this model might appear sufficient to describe the
long-term evolution of a Rydberg wave packet
in an alkali-metal atom.
However,
this is incorrect,
as we demonstrate next.

The long-term revival structure of hydrogenic wave packets
has already been studied for the case where the
laser excites a mean energy corresponding to $N^\ast = {\bar N}$,
where $\bar N$ is an integer
\cite{sr}.
Using a Taylor expansion of the energy
shows that the evolution of the packet
for times up to a time $t_{\rm sr}$ is governed by the
first three terms in the expansion.
At present,
we are interested in the general case with noninteger $N^\ast$.
The desired results can be obtained from the earlier
analysis by a judicious choice of variables.
Let us write $N^\ast = \bar N - \mu /\nu$,
where $\bar N$ is the smallest integer greater
than or equal to $N^\ast$
and ${\mu}/{\nu}$ is an irreducible fraction less than one.
In the present case of hydrogen,
this fraction represents the laser detuning
away from the nearby atomic resonance.

Expanding the hydrogenic energies to third order gives
\beq
E_n \simeq E_{N^\ast} + E_{N^\ast}^\prime (n - {N^\ast})
+ \fr 1 2 E_{N^\ast}^{\prime\prime} (n - {N^\ast})^2
+ \fr 1 6 E_{N^\ast}^{\prime\prime\prime} (n - {N^\ast})^3
\quad .
\label{expans}
\eeq
This expansion in powers of the noninteger quantity $(n-N^\ast)$
defines the time scales $T_{\rm cl} = 2 \pi N^{\ast 3}$,
$t_{\rm rev} = \fr 23 N^\ast T_{\rm cl}$,
and $t_{\rm sr} = \fr 34 N^\ast t_{\rm rev}$.
Introduce the integer quantity $k=(n-{\bar N})$
and define the new scales
\beq
T_{\rm cl}^\prime = T_{\rm cl}
\left( 1 - \fr {2 \mu} {\nu} \fr {T_{\rm cl}} {t_{\rm rev}}
+ \fr {3 \mu^2} {\nu^2} \fr {T_{\rm cl}} {t_{\rm sr}}
\right)^{-1}
\quad ,
\label{tclpr}
\eeq
\beq
t_{\rm rev}^\prime = t_{\rm rev}
\left( 1 - \fr {3 \mu} {\nu} \fr {t_{\rm rev}} {t_{\rm sr}}
\right)^{-1}
\quad ,
\label{trevpr}
\eeq
\beq
t_{\rm sr}^\prime = t_{\rm sr}
\quad ,
\label{tsrpr}
\eeq
in terms of which $\Psi (\vec r,t)$ may be written in the form
\beq
\Psi ({\vec r},t) = \sum_{k=-\infty}^{\infty} c_k
\varphi_k ({\vec r}) \exp \left[ -2 \pi i
\left( \fr {kt} {T_{\rm cl}^\prime}
-  \fr {k^2 t} {t_{\rm rev}^\prime}
+ \fr {k^3 t} {t_{\rm sr}^\prime} \right) \right]
\quad .
\label{hpsi3pr}
\eeq
Since by assumption $\bar N \gg 1$,
the lower limit in $k$ is well approximated by $-\infty$.
An overall complex phase has been dropped.

With these variable changes,
the wave function in
Eq.\ \rf{hpsi3pr}
has a structure similar to that used in the earlier analysis
with integer $N^\ast$.
However,
the three key time scales are modified
by amounts depending on the laser detuning in hydrogen,
i.e., the irreducible fraction $\mu/\nu$.
Following the approach of ref.\ \cite{sr},
we can obtain the times of formation $t_{\rm frac}^\prime$
of the subsidiary wave packets
and their periodicities $T_{\rm frac}^\prime$.
Write ${\bar N} = 4 \et + \la$,
where $\et$ and $\la$ are integers and $\la = 0,1,2$, or 3.
We then find
\beq
t_{\rm frac}^\prime = \fr 1 q t_{\rm sr}^\prime
- \fr m n t_{\rm rev}^\prime~~,~~~~
T_{\rm frac}^\prime = \fr 3 q t_{\rm rev}^\prime
- \fr u v T_{\rm cl}^\prime
\quad ,
\label{Thyd}
\eeq
where $q$ is an integer multiple of three
and
\beq
\fr m n=\fr 3 {4q}
\bigl(\la  - 5 \fr \mu \nu \bigr)\quad {\rm (mod~1)}~~,~~~~
\fr u v=\fr {2 (\eta + \la) - 3} {q}\quad {\rm (mod~1)}
\quad .
\label{fhyd}
\eeq
As expected,
when $\mu \rightarrow 0$
the expressions reduce to those of ref.\ \cite{sr},
corresponding to excitation on resonance.

These results show that,
using a hydrogenic expansion,
the times $t_{\rm frac}^\prime$ and
$T_{\rm frac}^\prime$ are completely determined
once $N^\ast$ has been fixed.
However,
this determination has come before specifying
the alkali-metal atom in question,
whereas quantum defects
are known to cause additional shifts in the revival times
\cite{sr2}.
We have therefore shown that describing a wave packet
in an alkali-metal atom purely with
hydrogenic energies and wave functions is insufficient
for a complete treatment.

We next turn to a different theoretical description
that does provide a more complete description
and therefore also permits an estimate of
the deviations from hydrogenic behavior.
Moreover,
it allows a quantitative comparison of the effects
of laser detuning and quantum defects.

Non-hydrogenic features of radial packets
in alkali-metal atoms can be incorporated analytically
via a supersymmetry-based quantum-defect theory
(SQDT),
which has analytical wave functions
with the asymptotic Rydberg series as exact energy eigenvalues
\cite{sqdt,susyreview}.
Since the SQDT wave functions
$\varphi_{\ast {n^\ast}} ({\vec r})$
both incorporate quantum defects
and form a complete and orthonormal set,
they can be used as a basis for an expansion of
a packet in an alkali-metal atom.
We write
\beq
\Psi ({\vec r},t) = \sum_{n^\ast} c_{n^\ast}
\varphi_{\ast {n^\ast}} ({\vec r})
\exp \left( -i E_{n^\ast} t \right)
\quad ,
\label{wavesqdt}
\eeq
where $c_{n^\ast}$ is a distribution in $n^\ast$ centered
on $N^\ast$.

Expanding the energy $E_{n^\ast}$ as before
around $N^\ast = {\bar N} - {\mu}/{\nu}$
defines the same three time scales as
for the hydrogenic expansion:
$T_{\rm cl} = 2 \pi N^{\ast 3}$,
$t_{\rm rev} = \fr 23 N^\ast T_{\rm cl}$,
and $t_{\rm sr} = \fr 34 N^\ast t_{\rm rev}$.
However,
in this case the expansion is in powers of
$(n^\ast - N^\ast)\equiv (k + {\al}/{\be})$,
where we have introduced
${\al}/{\be} = {\mu}/{\nu} - \de (l)$ (mod 1),
and $k$ is the integer part of $(n^\ast - N^\ast)$.
For excitation in alkali-metal atoms,
the laser detuning from the nearby atomic resonance is
$\al /\be$.
Excitation on resonance corresponds to $\al = 0$.
The quantity $\mu /\nu$,
which for the above analysis in hydrogen represented
the laser detuning,
is now instead the fractional part of the
sum of the laser detuning and the quantum defect.

Keeping the first three terms in the energy expansion gives
\beq
\Psi ({\vec r},t) = \sum_{k=-\infty}^{\infty} c_k
\varphi_{\ast k} ({\vec r}) \exp \left[ -2 \pi i
\left( \fr {kt} {T_{\rm cl}^{\ast \prime}}
-  \fr {k^2 t} {t_{\rm rev}^{\ast \prime}}
+ \fr {k^3 t} {t_{\rm sr}^{\ast \prime}} \right) \right]
\quad ,
\label{sqdtpsi3}
\eeq
where
\beq
T_{\rm cl}^{\ast \prime} = T_{\rm cl}
\left( 1 - \fr {2 \al} {\be} \fr {T_{\rm cl}} {t_{\rm rev}}
+ \fr {3 \al^2} {\be^2} \fr {T_{\rm cl}} {t_{\rm sr}}
\right)^{-1}
\quad ,
\label{sqdttclpr}
\eeq
\beq
t_{\rm rev}^{\ast \prime} = t_{\rm rev}
\left( 1 - \fr {3 \al} {\be} \fr {t_{\rm rev}} {t_{\rm sr}}
\right)^{-1}
\quad ,
\label{sqdttrevpr}
\eeq
\beq
t_{\rm sr}^{\ast \prime} = t_{\rm sr}
\quad .
\label{sqdttsrpr}
\eeq
It is important to note that these expressions depend on
both the quantum defect $\de$ and the laser detuning $\al /\be$.
Part of the dependence on $\al /\be$ is explicit.
The remainder and the dependence on $\de$ appears through
the hidden dependence on the irreducible fraction $\mu/\nu$,
which enters through the definitions
of the time scales in terms of $N^\ast$.

Following the analysis in
ref.\ \cite{sr2}
with ${\bar N} = 4 \et + \la$ as before,
we find that the packet may be written
as a sum of distinct subsidiary waves
at times $t_{\rm frac}^{\ast \prime}$
with periodicities $T_{\rm frac}^{\ast \prime}$
given by
\beq
t_{\rm frac}^{\ast \prime} = \fr 1 q t_{\rm sr}^{\ast \prime}
- \fr m n t_{\rm rev}^{\ast \prime}~~,~~~~
T_{\rm frac}^{\ast \prime} = \fr 3 q t_{\rm rev}^{\ast \prime}
- \fr u v T_{\rm cl}^{\ast \prime}
\quad ,
\label{Tsqdt}
\eeq
where
\beq
\fr m n = \fr {3} {4q} \left( \la - \fr {\mu} {\nu}
- \fr {4 \al} {\be} \right) {\rm (mod~1)}~~,~~~~
\fr u v = \fr {2 (\eta + \la) - 3} {q}
+ \fr 2 q \left( \fr {\al} {\be}
- \fr {\mu} {\nu} \right) {\rm (mod~1)}
\quad .
\label{fsqdt}
\eeq
The revival times and periodicities depend on the quantum defects.
Unlike in hydrogen,
specifying $N^\ast$ fails to determine
$t_{\rm frac}^{\ast \prime}$ and $T_{\rm frac}^{\ast \prime}$.
In addition,
the alkali-metal atom and the corresponding quantum defects must
be specified.
Moreover,
the scales
$t_{\rm frac}^{\ast \prime}$ and $T_{\rm frac}^{\ast \prime}$
in the present case evidently cannot be obtained from the
corresponding hydrogenic scales
$t_{\rm frac}^{\prime}$ and $T_{\rm frac}^{\prime}$
by a simple renormalization.

Equation \rf{fsqdt} also demonstrates that
the effects of quantum defects
are different from those of the laser detuning.
Replacing $\mu /\nu$ with its definition
in terms of the laser detuning $\al /\be$ and the quantum defect $\de$
shows that the time $t_{\rm frac}^{\ast \prime}$
and the period $T_{\rm frac}^{\ast \prime}$
have different dependences on $\de$ and $\al /\be$.
Therefore, laser detuning cannot
be mimicked by quantum defects or vice versa.
The differences arise because a constant shift
in the laser detuning is equivalent to a constant (opposite) shift
for all energy levels,
whereas a constant shift in the quantum defect would correspond
to varying shifts among the energy levels
since $E_{n^\ast} = 1/n^{\ast 2}$.
Furthermore,
these differences are of the order of
$t_{\rm rev}^{\ast \prime}$,
representing many classical orbital periods.

Next, we consider whether the
modifications of the long-term revival behavior
are experimentally observable with current technology.
For simplicity in what follows,
we take an example for which the excitation is at resonance.
Consider a wave packet in rubidium with $N^\ast \simeq 45.35$,
which is of the type that can readily be produced experimentally.
For definiteness set $q=6$,
which would produce a large peak in
the ionization signal in a pump-probe experiment
as it corresponds to the formation of a
single packet more closely resembling the initial packet
than does the original full revival.
We then find ${\bar N} = 46$, $\et =11$, $\la = 2$,
$\mu = 13$, and $\nu = 20$.
The description of long-term revivals using
hydrogenic energies and wave functions
yields $t_{\rm frac}^{\prime} \simeq 2.05$ nsec,
with a periodicity
$T_{\rm frac}^{\prime} \simeq 215$ psec.
In contrast,
the description incorporating quantum defects
also requires a specification of the quantum defect $\de (1)$ for
p states of rubidium,
which we take to be $\de (1) \simeq 2.65$.
This gives ${\al}/{\be} = 0$
and $t_{\rm frac}^{\ast \prime} \simeq 2.36$ nsec,
with periodicity
$T_{\rm frac}^{\ast \prime} \simeq 206$ psec.

This analysis shows that there is a difference of approximately
$0.31$ nsec between the predicted times for the
occurrence of the $q=6$ long-term revival.
This is greater than $20$ classical orbital periods.
A discrepancy of this size should be measurable
in a pump-probe experiment with a delay line
of approximately $2.5$ nsec.

Additional support for the above analysis
comes from the numerical computation of $t_{\rm frac}^{\ast \prime}$
and $T_{\rm frac}^{\ast \prime}$.
The initial packet is taken as a p-state radial
squeezed state for rubidium,
with mean energy and mean radius at the outer apsidal
point determined by the value $N^\ast \simeq 45.35$.
We numerically evolve this packet
using the time-dependent Schr\"odinger radial equation
with the SQDT effective potential.

Figure 1 shows the absolute square of the autocorrrelation
function for the radial squeezed states for times
up to 3 nsec.
For this case,
$t_{\rm rev} \simeq 0.43$ nsec.
The fractional revivals at $t \simeq 0.07$ nsec,
$0.11$ nsec, and $0.22$ nsec agree with
the predicted values $\fr 1 6 t_{\rm rev}$,
$\fr 1 4 t_{\rm rev}$, and $\fr 1 2 t_{\rm rev}$,
respectively.
The peaks in the autocorrelation function
at the full revival time $t_{\rm rev}$
are diminished relative to those at
$\fr 12 t_{\rm rev}$,
due to distortions of the packet arising from
the higher-order terms in the energy expansion.
Long-term revivals with $q = 12$, $9$, and $6$ are visible,
at times in good agreement with the predictions
$t_{\rm frac}^{\ast \prime} \simeq 1.17$ nsec, $1.57$ nsec,
and $2.36$, respectively.
Furthermore,
the peaks have periods agreeing with
the predicted values
$T_{\rm frac}^{\ast \prime} \simeq 0.10$ nsec,
$0.14$ nsec, and $0.21$ nsec for $q = 12$, $9$, and $6$.

Figure 2 displays the Fourier transform of the numerical data for
the autocorrelation function shown in Fig.\ 1.
The dominant frequencies agree
with those obtained from the energy spacings,
$\nu = \De E_{n^\ast}/2 \pi $,
with $E_{n^\ast} = - 1/{2 n^{\ast 2}}$.
The beating of these frequencies
leads to the revival structure observed in Fig.\ 1.
This confirms that the
Schr\"odinger equation with the SQDT effective potential
generates the correct eigenenergies for an alkali-metal atom.

Analytically,
we find $\De r \De p_r \simeq 0.503$ for the
packet at the outer apsidal point.
The uncertainty product then oscillates as a function of time.
Near $t_{\rm frac}^{\ast \prime}$ with $q=6$,
$\De r \De p_r$ is less than its value near $t_{\rm rev}$,
indicating that the $q=6$ long-term revival
is closer to minimum uncertainty than the full revival.
Figure 3 shows the uncertainty ratio $\De r /\De p_r$
as a function of time.
The squeezing of the wave packet is
visibly more extreme near the $q=6$ long-term revival at
$t \simeq 2.36$ nsec than at the full revival
near $t \simeq 0.43$ nsec.
The uncertainty ratio oscillates around a mean value
close to 50000 in atomic units.
This agrees with the value of
$\De r /\De p_r$ calculated analytically
in SQDT for an energy eigenstate with $n^\ast \simeq 45.35$.
Evidently,
the long-term uncertainty ratio takes this mean value.

The analytical and numerical results we have presented
in this work show that modifications of the hydrogenic
long-term revival structure arise
at an experimentally observable level
for radial Rydberg wave packets in alkali-metal atoms.
A description purely in terms of hydrogenic wave
functions and energies is insufficient for an accurate
prediction of the revival times and periodicities.
Moreover,
the effects of laser detuning and the quantum defects
are different.
An analytical treatment is possible within the context of SQDT.
The predicted values of
$t_{\rm frac}^{\ast \prime}$ and $T_{\rm frac}^{\ast \prime}$
for rubidium agree with those observed in the long-term
behavior of a radial squeezed state.

\vfill\eject

\begin{description}

\item[{\rm Fig.\ 1:}]
Absolute square of the autocorrelation function for
rubidium radial squeezed states with $N^\ast \simeq 45.35$
as a function of time in nanoseconds.

\item[{\rm Fig.\ 2:}]
Fourier spectrum for the absolute square of the
autocorrelation function
of rubidium radial squeezed states
with $N^\ast \simeq 45.35$.
The unnormalized Fourier transform is plotted as
a function of frequency in inverse nanoseconds.

\item[{\rm Fig.\ 3:}]
The ratio of uncertainties $\De r / \De p_r$ in
units of $10^5$ a.u.\ as a function of time in
nanoseconds for a radial squeezed state of rubidium
with $N^\ast \simeq 45.35$.

\end{description}

\vfill\eject


\begin{thebibliography}{xx}
\bibitem{ftnt}
Note that Ref.\ \cite{ps} also presents
a long-term revival in hydrogen at $t= \bar n^3 T_{\rm cl}$.
\end{thebibliography}
\end{document}